\documentclass[]{pasj02} 
\usepackage[switch,mathlines]{lineno} 
\usepackage{natbib}
\usepackage[version=4]{mhchem}
\newcommand{\citeay}[1]{\citeauthor{#1} \citeyear{#1}}

\jyear{2025}
\Received{}
\Accepted{}

\begin{document} 

\title{
Possible time-variable iron-K$\alpha$ emission in the circumnuclear region of the Circinus galaxy
}

\author{
Aiko \textsc{Miyamoto},\altaffilmark{1}\altemailmark\orcid{0009-0009-2127-8830} \email{miyamoto@ess.sci.osaka-u.ac.jp} 
 Taiki \textsc{Kawamuro},\altaffilmark{1}\orcid{0000-0002-6808-2052}
 Hirokazu \textsc{Odaka},\altaffilmark{1,2}\orcid{0000-0003-2670-6936}
 Takuma \textsc{Izumi},\altaffilmark{3}\orcid{0000-0001-9452-0813}
 and 
 Hironori \textsc{Matsumoto}\altaffilmark{1,2}
}
\altaffiltext{1}{Department of Earth and Space Science, Graduate School of Science, The University of Osaka, 1-1 Machikaneyama, Toyonaka, Osaka 560-0043, Japan}
\altaffiltext{2}{Forefront Research Center, Graduate School of Science, The University of Osaka, 1-1 Machikaneyama, Toyonaka, Osaka 560-0043, Japan}
\altaffiltext{3}{National Astronomical Observatory of Japan, 2-21-1 Osawa, Mitaka, Tokyo 181-8588, Japan}

\KeyWords{galaxies: active --- galaxies: individual (Circinus) --- X-rays: galaxies}  

\maketitle

\begin{abstract}
We present imaging and spatially resolved spectral analyses of eight Chandra data taken for the Circinus galaxy in $\approx$ 22 years to reveal neutral iron-K$\alpha$ emission on a circumnuclear scale ($\sim$ 10--100\>pc) and search for time variability in the emission.
By simulating and taking account of point-source emission from the active galactic nucleus (AGN), we detect iron-line emission 
$\sim$ 20--60 pc away from the nucleus, particularly in the eastern and western regions. 
In the two regions, possible time variability in the line flux was also detected. 
Our spectral analysis then finds that the observed equivalent widths can reach $\sim$ 2\>keV and the slopes of underlying continua are rather inverted with $\Gamma < 0$. 
These are consistent with a scenario in which the iron emission originates from clouds illuminated by AGN X-rays; 
our result could provide the first extragalactic example of AGN X-ray echoes. 
In this scenario, we estimated the physical sizes of the illuminated clouds based on the timescale of variability to be less than 6 pc. 
Furthermore, we compared the iron emission distribution with the cold molecular distribution inferred by Atacama Large Millimeter/submillimeter Array (ALMA) observation of CO($J$=3--2), revealing that in the region of bright iron-line emission, the molecular emission seems to be weak. 
This might suggest that the AGN X-ray emission affects the chemical composition in the form of AGN feedback. 
\end{abstract}


\section{Introduction}

Understanding the distribution and kinematics of gas around active galactic nuclei (AGNs) is essential for discussing supermassive black hole (SMBH) growth and feedback. For instance, observations of cold molecular gas emission are invaluable for tracing cold material and have provided significant insights into the gas properties around AGNs at various spatial scales from a few 100 pc to sub parsec (e.g., \citeay{Elmouttie1998}; \citeay{Alonso-Herrero2020}; \citeay{Izumi2018}, \citeyear{Izumi2023}). However, molecular lines can be suppressed by the dissociation or heating of molecules due to intense radiation from the central engine, potentially leading to incomplete views of the environment. 
In such situations, the neutral iron-K$\alpha$ emission line at 6.4\>keV can be a complementary probe. 
It arises from material illuminated by hard X-rays and is relatively insensitive to the chemical state of the gas (i.e., atomic, molecular, and partially ionized states). 
As a result, the Fe-K$\alpha$ line allows us to reveal the matter that may be missed by molecular gas observations, making it indispensable for obtaining a comprehensive view of the nuclear environment.

High-spatial-resolution X-ray observations by Chandra have revealed that neutral Fe K$\alpha$ emission spatially extends over tens to hundreds of parsecs in nearby AGNs: 
NGC 1068 (\citeay{Young2001}; \citeay{Bauer2015}; \citeay{Nakata2021}),  NGC 4945 (\citeay{Marinucci2012}, \citeyear{Marinucci2017}), 
ESO 428-G014 (\citeay{Fabbiano2017}), 
NGC 5643 (\citeay{Fabbiano2018}), 
NGC 2110 (\citeay{Kawamuro2020}), 
MKN 573, NGC 1386, NGC 3393, NGC 5643, NGC 7212 (\citeay{Jones2021}), 
and the Circinus galaxy (\citeay{Marinucci2013}; \citeay{Kawamuro2019}).
These results indicated that X-ray reprocessing can occur over a wide range of spatial scales. 
Furthermore, by comparing molecular gas and Fe-K$\alpha$ line distributions, some studies showed that they are spatially separated (e.g., \citeay{Feruglio2020}; \citeay{Kawamuro2021}). 
This suggests that X-ray irradiation or AGN emission may alter the chemical state of the surrounding gas in ways that molecular lines cannot fully trace.

Despite the above accumulated amount of observational studies, the temporal evolution of the extended Fe-K$\alpha$ emission induced by changes in irradiating AGN luminosity has not been explored. 
An exception is the probable X-ray echo in our Galaxy due to Sagittarius A$^\ast$ (hereafter, Sgr A$^\ast$; \citeay{Koyama2018}). 
Investigating such time variability can potentially constrain the three-dimensional distribution of gas clouds and, more simply, the size of an irradiated region at resolutions better than those of Chandra. 
The size constraint is indeed valuable for robustly comparing the irradiated region with high-spatial-resolution images at different wavelengths ($<$ 1\arcsec) 
and allows us to discuss the interaction between AGN radiation and interstellar medium more accurately. 

In this paper, we investigated the distribution of the matter in the Circinus galaxy (hereafter, Circinus) on a circumnuclear scale using Chandra with a particular focus on the time variation of iron line flux. 
Circinus with a redshift of $z = 0.00145$ is located at 
a distance of 4.2 Mpc (\citeay{Freeman1977}) and hosts one of the nearest Compton-thick AGNs 
(\citeay{Matt1999}; \citeay{Soldi2005}; \citeay{Yang2009}; \citeay{Arevalo2014}).
The short distance allows us to achieve high physical resolution (i.e., 20 pc/1\arcsec). 
The Compton-thick obscuration in the line of sight prevents the central AGN X-ray emission from outshining ambient reflected X-ray emission, including iron-K$\alpha$ one. 
Thus, Circinus is an ideal target to spatially resolve and discuss X-ray irradiated gas. 
In the past, \citet{Marinucci2013} and \citet{Kawamuro2019} investigated spatially extended iron-K$\alpha$ emission using Chandra, but they did not discuss any time variability. 
Thus, our study is unique with respect to the previous ones. 

The remainder of this paper is constructed as follows. Section~\ref{sec:obsdata} presents the reduction of Chandra observation data. Section~\ref{sec:data_analysis} provides an overview of the data analysis. Sections~\ref{sec:discussion} and ~\ref{sec:conc} present our discussions and conclusions, respectively. Unless otherwise noted, the errors are quoted at a 90\% confidence level for a single parameter of interest.

\section{Observations and standard data reduction}\label{sec:obsdata}

Circinus was observed 14 times on-axis by Chandra with the Advanced CCD Imaging Spectrometer (ACIS: \citeay{Garmire03}) from 2000 to 2022. 
The on-axis data are crucial, as they provide us with images at the highest resolutions around the galaxy achievable in the X-ray band.
Among them, we utilized  eight observations whose exposure times are more than 20 ks, to retain good photon statistics. 
Table~\ref{tab:observations} lists the eight selected observations, including the data in 2010 used in a previous study (\citeay{Marinucci2013}; \citeay{Kawamuro2019}). 
The eight data left still make it possible to investigate time variation in spatially extended Fe-K$\alpha$ emission across 22 years. 
Of the eight observations, five used the High-Energy Transmission Grating (HETG: \citeay{Canizares05}). 

For the standard data reduction, 
we used  Chandra Interactive Analysis of Observations (CIAO) 4.16.2 with the Chandra Calibration Data Base 4.11.5. 
We reprocessed the raw data taken in the eight observations using the CIAO tool \texttt{chandra\_repro} to produce a cleaned event file for each observation. 
Then, to identify the observation periods when the background count abruptly increased, we created a light curve for each observation in the 0.5--10.0\>keV band in an annular region from  0\farcm8 to 1\farcm1 centered at the AGN. In the region, bright sources are absent, and the light curve should reflect background count rates. 
Eventually, no clear flares were found, and no data were excluded due to background flares.


\begin{table}[t]
  \tbl{Chandra observation data used in this study.\footnotemark[$*$] }{%
  \begin{tabular}{cccc}
      \hline\noalign{\vskip2pt}
      ObsID & Date & Exposure (ks) & Grating  \\   [1pt]
      (1) & (2) & (3) & (4) \\   [1pt]
      \hline\noalign{\vskip2pt}
      356 & 2000/3/14 & 24.72 & NONE \\   [2pt] 
      62877 & 2000/6/16 & 60.22 & HETG \\   [2pt]
      4770 & 2004/6/2 & 50.03 & HETG \\   [2pt]
      4771 & 2004/11/28 & 58.97 & HETG \\   [2pt]
      12823 & 2010/12/17 & 152.36 & NONE \\   [2pt]
      12824 & 2010/12/24 & 38.89 & NONE \\   [2pt]
      26037 & 2022/7/22 & 40.08 & HETG \\   [2pt]
      26091 & 2022/7/28 & 30.08 & HETG \\   [2pt]
      \hline\noalign{\vskip2pt}
    \end{tabular}}\label{tab:observations}
\begin{tabnote}
\footnotemark[$*$] Columns: (1) Observation ID. (2) Observation start date. (3) Exposure time. 
(4) Verification of the grating setting.  \\
\end{tabnote}
\end{table}


\section{Data analyses and results}\label{sec:data_analysis}

\subsection{Flow of the analyses}\label{ssec:flowchart}

Our interest is in the extended emission around the unresolved nucleus.
To reveal the emission, we need to consider that even the point-like nuclear emission extends due to the finite spatial resolution, or the point-spread function (PSF), and contaminates outer in-situ emission.

To reveal the outer in-situ emission, we took several steps. 
First, while considering pileup effects, we determined a spectral model of the nuclear emission in each observation (subsection~\ref{ssec:nuemi}). 
Secondly, considering the emission solely from the nucleus that has the determined model, we simulate a Chandra image (sub-subsection~\ref{sssec:nusim}).  
Thirdly, we took the ratio of the observed image to the simulated one; 
the ratio image enables us to discuss in-situ extended emission (subsection~\ref{ssec:ironmap}). 
Finally, to derive physical quantities important to discuss emitting sources, or mechanisms (e.g., Fe-K$\alpha$ flux and its equivalent width), we constrained the spectra of the extended emission while taking account of the contaminating emission from the nucleus  (subsection~\ref{ssec:cnrspec}).

\subsection{Emission from the nucleus}\label{ssec:nuemi} 

\subsubsection{Pileup}\label{sssec:pileup} 

The nucleus of Circinus is bright enough to cause photon pileup in counting photons with Chandra. 
The effect needs to be considered to estimate the true spectrum from the observed one, and the true one is indispensable to simulate Chandra observations. 

We examined whether the pileup effect was significant in each observation using the $\mathtt{pileup\_map}$ script. 
The script computes the number of counts per ACIS frame at each CCD pixel. The counts per frame can be then used to estimate the fraction of piled events to the total detected events\footnote{$\langle$https://cxc.harvard.edu/csc/memos/files/Davis\_pileup.pdf$\rangle$.}. 
We then estimated the pileup fraction from the counts per frame at each pixel by referring to a conversion factor\footnote{$\langle$https://cxc.cfa.harvard.edu/ciao/ahelp/pileup\_map.html$\rangle$.}. 
In the three observations without HETG (ObsID = 356, 12823 and 12824), the maximum pileup fraction reached $\approx$ 27\% in the nuclear region within $\approx$ 1\arcsec.
In the other observations with HETG, even the highest fraction was well below 10\% and the pileup effect can be negligible for them.
For further analyses, we considered the pileup effect only for the three observations without HETG, where the highest pileup fraction exceeded 10\%.

\subsubsection{Nuclear spectra}\label{sssec:nuspec} 

To extract nuclear spectra, 
we determined the detector coordinates of the nucleus for each observation. 
We created X-ray count images with a pixel size of $0.0615\times0.0615\>\mathrm{arcsec^{2}}$ using energy-dependent subpixel event repositioning (e.g., \citeay{Tsunemi2001}; \citeay{Li2003}, \citeyear{Li2004}). This process creates images at an angular resolution higher than the Chandra CCD pixel size of 0\farcs492.
Based on the sub-pixel count image, we identified the AGN count peak and created two histograms 
by projecting the region of the square inside a 0\farcs5 radius circle from that peak along the X and Y coordinates in each image.
The 0\farcs5 radius was adopted so that the nuclear emission was dominant.
Then, we fit each of the resultant histograms with the sum of the Gaussian and constant components 
and defined the mean values of the Gaussian components as the nuclear coordinates.

For each observation, we extracted a source spectrum from a 1\arcsec circle and a background spectrum from an annulus from 0\farcm8 to 1\farcm1; both regions were centered at the above determined coordinates. We note that, for the five HETG observations, 
only their 0th order spectra were extracted and used. 
In the background region, there were no obvious bright sources.  
The response files, appropriate for a point source, were generated using the CIAO tool $\mathtt{specextract}$. 
The spectra were then binned so that each bin retained at least one count. 
The best-fit parameters were determined using the C-statistic technique in XSPEC Version 12.14.0h (\citeay{Arnaud1996}).

To fit the nuclear spectra in the 0.5--8.0\>keV band, we adopted a phenomenological model. 
The phenomenological one is sufficient as, in estimating how largely the nuclear emission contaminates the outer in-situ emission, merely the spectral shape is important. The actual model is 
\begin{equation}
\mathtt{TBabs1 \times ( TBabs2 \times powerlaw1 + powerlaw2 + emission~lines )}, 
\end{equation}
in the XSPEC terminology. 
The \texttt{TBabs1} and \texttt{TBabs2} correspond to \texttt{TBabs} and consider the photoelectronic absorption by neutral gas. The \texttt{powerlaw1} and \texttt{powerlaw2} correspond to \texttt{powerlaw}, producing a simple power-law spectrum. 
\texttt{TBabs1} is ascribed to the Galactic absorption and the hydrogen column density was fixed at $N_{\rm H}$ = $5.30\times10^{21}\>\mathrm{atoms\>{cm}^{-2}}$, estimated by using the FTOOLS command $\mathtt{NH}$ with the coordinates of Circinus. 
In contrast, we used \texttt{TBabs2} together with \texttt{powerlaw1} to approximate a reflected continuum component. 
To reduce the number of free parameters, the photon index of \texttt{powerlaw1} was fixed at $\Gamma$ = 2.31 based on a past Circinus study by \citet{Arevalo2014}. 
We note that although \citet{Arevalo2014} included an absorbed power-law component in their spectral modeling, that component contributes negligibly in the 0.5--8\>keV band, whereas in our analysis the absorbed power law is used as a proxy for the reflection continuum. To verify that our approximation is valid, we replaced the absorbed power law with a \texttt{pexrav} reflection model with $\Gamma =1.6$ \citep{Arevalo2014} and carried out subsequent fits, confirming that our conclusions are essentially unchanged. This is physically natural, since below $\sim$\>10\>keV the reflection spectrum is largely shaped by photoelectric absorption, making it resemble that of an absorbed power law.
The \texttt{powerlaw2} model was used to reproduce remaining continuum emission, such as scattered continuum. Lastly, we applied a number of Gaussian components ($\mathtt{zgauss}$ in XSPEC) for emission lines. 
Emission lines were surveyed and identified by \citet{Sambruna2001} and \citet{Arevalo2014}, who used HETG spectra (ObsID = 374, 62877, 4770, 4771, 10226, 10223, 10832, 10833, 10224, 10844, 10225, 10842, 10843, 10873, 10872 and 10850). 
The energies of the identified lines were fixed based on \citet{Sambruna2001} and \citet{Arevalo2014}, and line widths were fixed at 0\>eV except for the Fe-K$\alpha$ line around 6.4\>keV. 
The energy and width of the Fe-K$\alpha$ line were allowed to vary freely in each observation. 
The other free parameters were the normalization of the two power-law components and the line normalizations for each observation. 

\begin{figure}[t]
 \begin{center}
  \includegraphics[width=8cm]{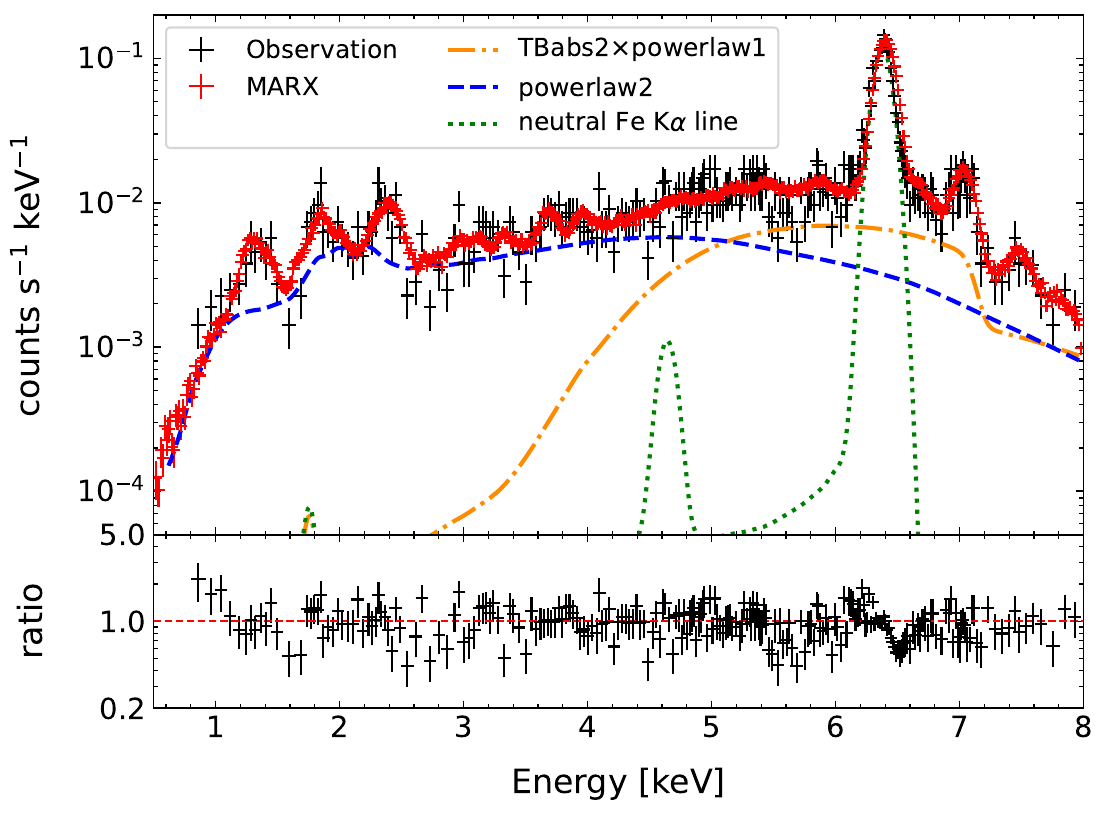} 
 \end{center}
\caption{Nuclear spectra: 0.5--8.0\>keV spectra of the observation (black) and MARX simulation (red) extracted from the central 1\farcs0 radius region on 2000 June 16. 
The orange dash-dot line, the blue dashed line, and the green dotted line the $\mathtt{TBabs2 \times powerlaw1}$ , $\mathtt{powerlaw2}$, and neutral Fe K$\alpha$ line components of the model shown in equation (1), respectively.
The ratio of the observation to MARX simulation is shown in the lower panel. 
}\label{fig:spec_obs_marx}
\end{figure}

As the spectra of the three non-grating observations severely suffer from the pileup effect particularly around the nucleus (sub-subsection~\ref{sssec:pileup}), 
we started by simultaneously fitting the other five spectra taken by the grating observations. 
This aim was to constrain some parameters of the spectral model and fix them across all spectra: specifically, 
the photon index of \texttt{powerlaw2} and 
the hydrogen column density of \texttt{TBabs2}. 
As a result of the fit, we obtained the photon index of 0.40 and $N_{\rm H}$ = $71.3\times10^{22}\>\mathrm{atoms\>{cm}^{-2}}$ ($C$-statistic$/$d.o.f. $= 2242.95/2117$). 
Figure~\ref{fig:spec_obs_marx} shows a fitting result for the spectrum obtained on 2000 June 16.
While fixing the above two parameters, we individually fitted all spectra with the phenomenological model. 
Particularly for the three spectra taken without HETG, 
we incorporated the CCD pileup model for Chandra into the above spectral model (sub-subsection~\ref{sssec:pileup}). 
The parameters of the pileup model were fixed to the values recommended 
for each observation by "The Chandra ABC Guide to Pileup"\footnote{$\langle$https://cxc.harvard.edu/ciao/download/doc/pileup\_abc.pdf$\rangle$.}, except for an $\alpha$ parameter. 
The $\alpha$ parameter represents the probability that the piled event is retained as an event with a good grade.


\begin{figure}[t]
 \begin{center}
  \includegraphics[width=\linewidth]{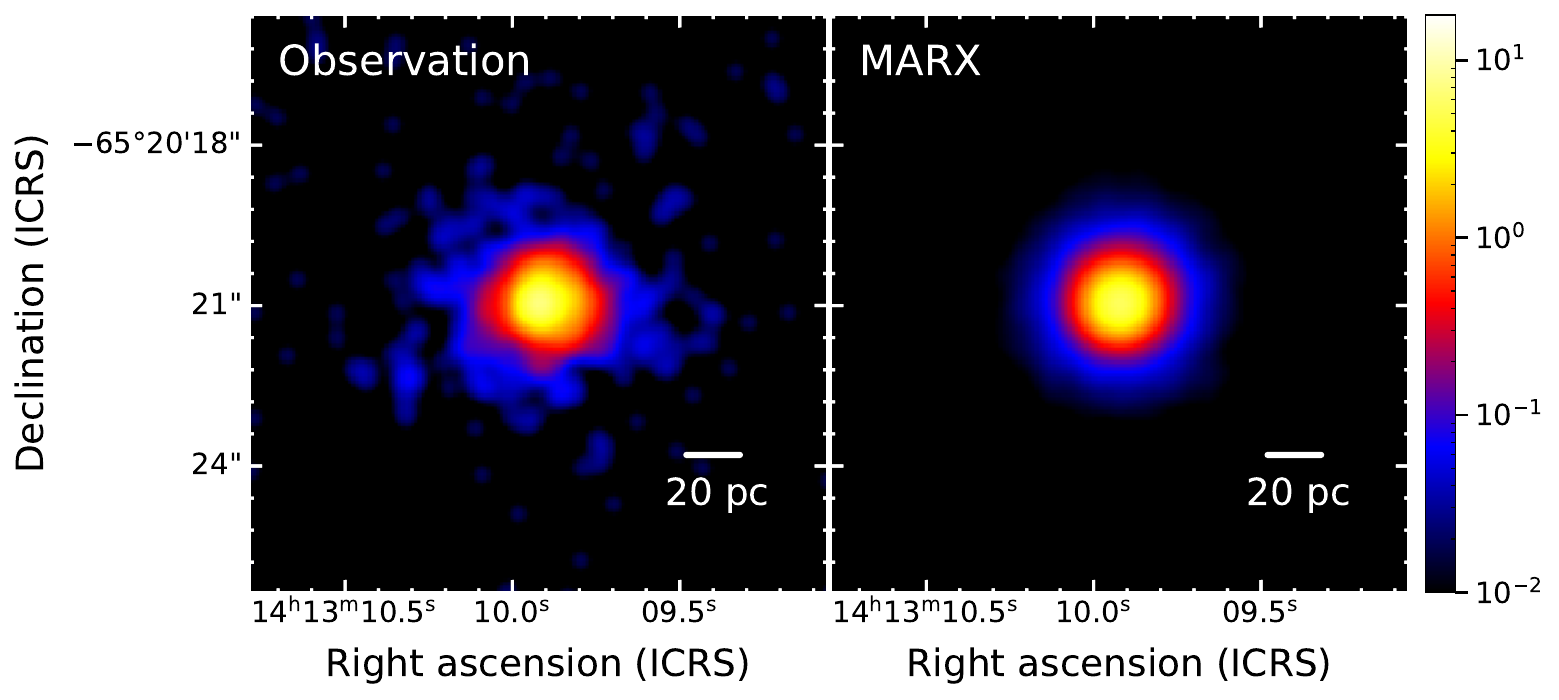} 
 \end{center}
\caption{Images of the observation (left) and MARX simulation (right) on 2000 June 16 in the 6.2--6.5\>keV band at 1/8 sub-pixel binning (smoothed with a Gaussian kernel of radius = 6\>pixel and $\sigma =$ 3\>pixel). Images were created on a logarithmic scale, with colors corresponding to the number of counts per pixel.
}\label{fig:img_obs_marx}
\end{figure}



\begin{figure*}[ht]
 \begin{center}
  \includegraphics[width=16cm]{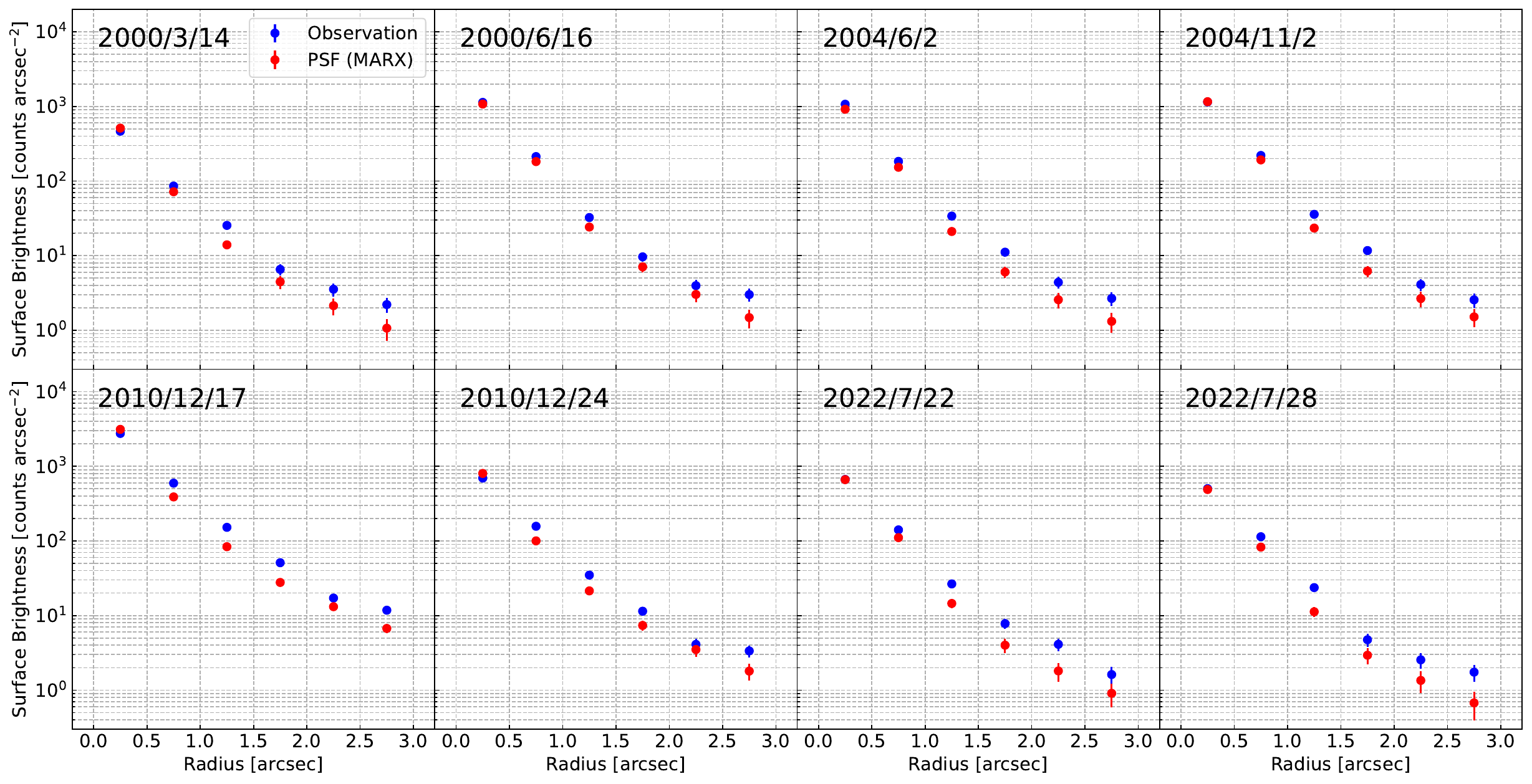} 
 \end{center}
\caption{Radial profiles of the observation (blue) and MARX simulation (red) in the 6.2--6.5\>keV band for each observation. 
}\label{fig:radial_profile}
\end{figure*}



\begin{figure*}[h]
 \begin{center}
  \includegraphics[width=\linewidth]{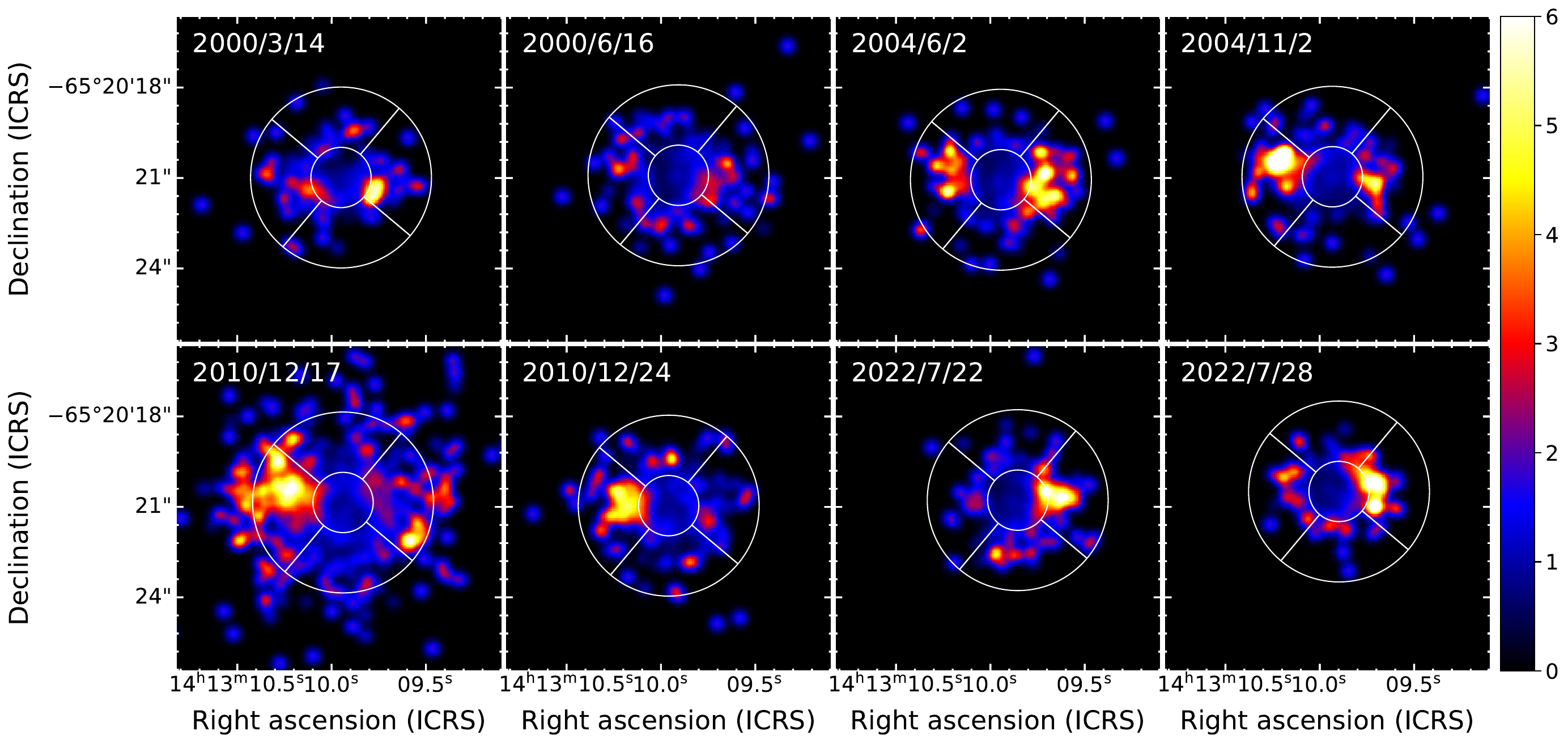} 
 \end{center}
\caption{ Neutral iron K$\alpha$ emission line maps for each observation. They were created in the 6.2--6.5\>keV band by dividing the number of counts per pixel in the observation image by those in MARX simulation image. 
Each pixel has a size of $0.0615\times0.0615\>\mathrm{arcsec^2}$, which is 1/8 of the original pixel size, and smoothing was performed with a Gaussian kernel of radius = 6\>pixel and $\sigma =$ 3\>pixel. 
The color bar represents the ratio of the observation to the simulation.
The inner 1\farcs0 radius circle and the outer 3\farcs0 radius circle define the region for extracting spectra in subsection~\ref{ssec:cnrspec}.
}\label{fig:Fe_map}
\end{figure*}



\begin{figure*}[ht]
 \begin{center}
  \includegraphics[width=16cm]{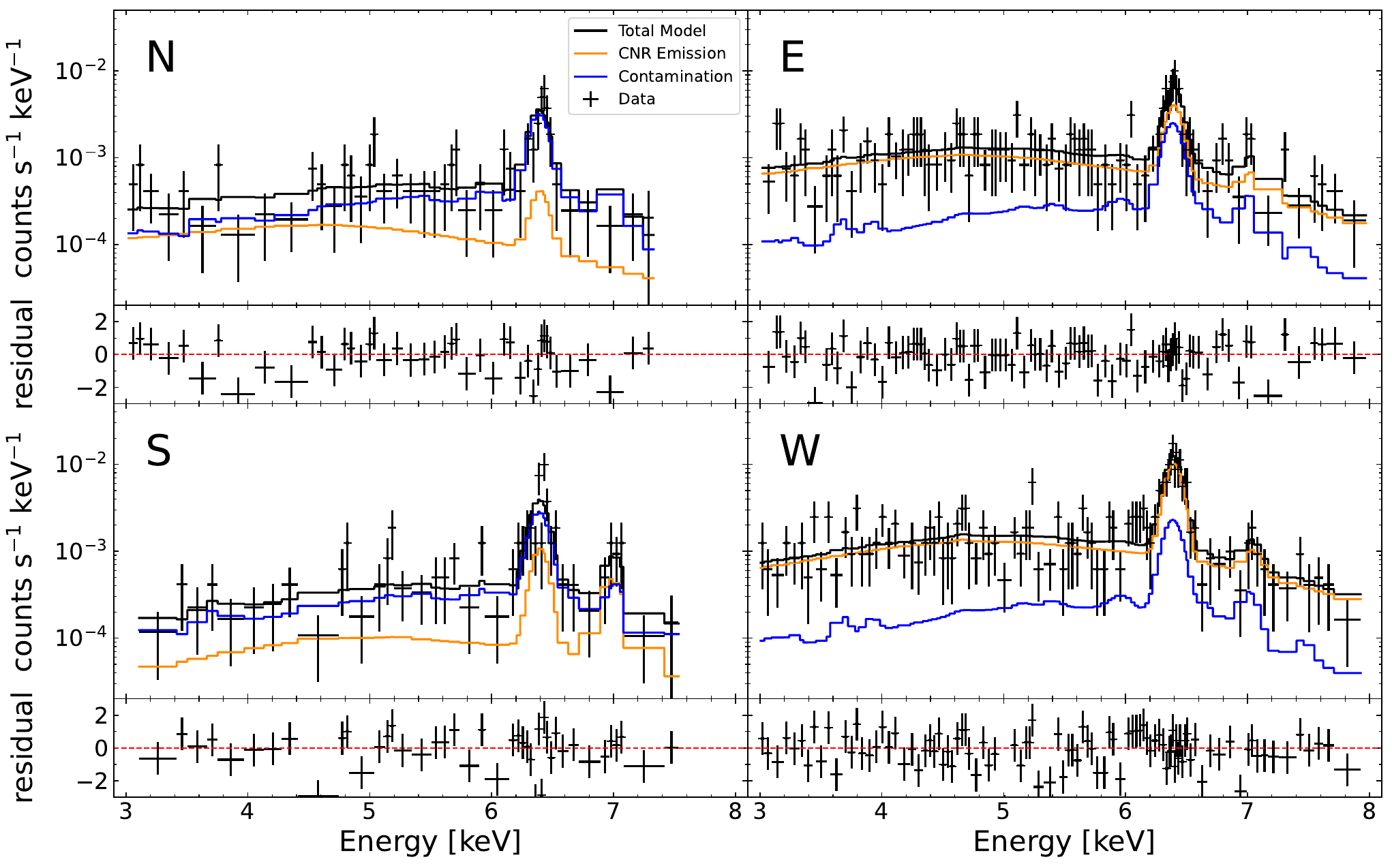} 
 \end{center}
\caption{ Spectra and models of the four circumnuclear regions (CNRs) on 2004 June 2; in-situ radiation from the CNR (orange), contamination from the center to the CNR (blue), and the sum of them (black).
}\label{fig:cnrspec}
\end{figure*}


\subsubsection{MARX simulation}\label{sssec:nusim}

We simulated each observation of a point source using the spectrum estimated in sub-subsection~\ref{sssec:nuspec}. 
We utilized a Monte Carlo-based ray-tracing simulator for Chandra observations of MARX \citep{Davis2012}, or the $\mathtt{simulate\_psf}$ command in CIAO. 
The version we used was 5.5.3. 
MARX can produce an event file expected from a Chandra observation, while taking account of various factors, including the mirror and detector responses, the satellite attitude, and a source spectrum. 
MARX needs an input event file, and the reprocessed event file was set. 
The other simulation parameters required are the central AGN coordinates, an additional amount of blurring to the PSF, and a switch for the pileup consideration. 
The coordinates were set to those determined in sub-subsection~\ref{sssec:nuspec}. 
Regarding the blurring, \citet{Ma2023} investigated suitable values of blur for ACIS-S observations and concluded that values larger than the default value 0\farcs07 are unlikely to be appropriate. We thus adopted 0\farcs07.
Lastly, the pileup was considered for all simulations; the effect was negligible for the grating observations. 

To reduce the statistical uncertainty of the simulation results, for each observation, we performed 100 MARX simulations with the same setup. Then, we took their average. 
For instance, figure~\ref{fig:spec_obs_marx} compares the observed and simulated spectra within the central 1\arcsec-radius circular region on 2000 June 16, indicating a good agreement. 
For the same observation day, the observed and simulated images are presented in figure~\ref{fig:img_obs_marx}. 
For ease of quantitative comparison, we summarize the radial profiles for all observation dates in figure~\ref{fig:radial_profile}. Generally, the profiles of the simulated data were lower than the observed ones, 
suggesting that there may be extended emission outside of 1\arcsec from the center.

\subsection{Mapping of iron K$\alpha$ emission lines}\label{ssec:ironmap}

For each observation, we created a 6.2--6.5\>keV image divided by the simulated image incorporating only nuclear emission (sub-subsection~\ref{sssec:nusim}); 
we refer to the image as the iron emission map afterward. 
Each ratio image was carefully constructed, while the peaks of the observed and simulated count images were matched at a pixel scale of $0.0615\times0.0615\>\mathrm{arcsec^{2}}$. 
Then, the resultant images were smoothed with a Gaussian kernel with $\sigma =$ 3 pixel. 
Figure~\ref{fig:Fe_map} shows the iron emission maps obtained. 
Some regions of bright emission appeared 20--60\>pc away from the AGN. 
Considering that the emitting regions seem to extend east and west, we divided the surrounding annulus of 20--60\>pc into four fan-like areas, labeled E, W, N and S, according to their orientation (figure~\ref{fig:Fe_map}). 
The regions E and W seem to not only show bright iron emission, but also vary the brightness across different epochs. 
The region E was brighter in 2004 and 2010 than in 2000, and became fainter again in 2022. 
The region W was brighter in 2004 June than in 2000, then became fainter in 2004 November and 2010, and then became brighter again in 2022. 
The other regions N and S seem to show almost no brightening of iron emission, unlike the regions W and E.


\begin{table*}[h]
  \tbl{Best-fit parameters.\footnotemark[$*$]}{%
  \begin{tabular}{ccccccc}
      \hline\noalign{\vskip2pt}
      Obs. Date & Region & $\Gamma$ & Norm. & $f_{\mathrm{FeK\alpha}}$ & ${\mathrm{EW}}_{\mathrm{FeK\alpha}}$ (keV) & C-stat./d.o.f. \\   [1pt]
      (1) & (2) & (3) & (4) & (5) & (6) & (7) \\   [1pt]
      \hline\noalign{\vskip2pt}
      2000/3/14 &	N &	$...$	&	$\dagger$ &	$<0.25$ &	$...$ &	$47/53$   \\   [2pt]
	  &	E &	$-0.11^{+0.79}_{-0.47}$ &	$7.0^{+8.2}_{-3.6}\times10^{-6}$ &	$0.75^{+0.35}_{-0.30}$ &	$0.9^{+0.8}_{-0.5}$ &	$176/207$   \\   [2pt]
	  &	S &	$<0.84$ &	$1.3^{+198.2}_{-0.9}\times10^{-8}$ &	$0.29^{+0.29}_{-0.24}$ & $...$ &		$83/84$		\\   [2pt]	
	  &	W &	$1.2^{+0.9}_{-0.6}$ &	$3.9^{+7.5}_{-2.6}\times10^{-5}$ &	$0.85^{+0.30}_{-0.34}$ &	$2.4^{+2.4}_{-1.2}$ &	$127/156$    \\   [2pt]
      \hline\noalign{\vskip2pt}
      2000/6/16 &	N &	$0.48^{+6.07}_{-2.95}$ &	$2.8^{+3303.3}_{-2.8}\times10^{-6}$ &	$<0.24$ &			$...$ &			$56/92$    \\   [2pt]
	  &	E &	$0.59^{+0.40}_{-0.40}$ &	$3.8^{+3.2}_{-1.7}\times10^{-5}$ &	$0.38^{+0.26}_{-0.23}$ &	$0.3^{+0.3}_{-0.2}$ &	$250/239$    \\   [2pt]
	  &	S &	$...$ &			$\dagger$ &				$0.27^{+0.23}_{-0.20}$ &	$...$ &			$61/66$    \\   [2pt]
	  &	W &	$-0.0053^{+0.5886}_{-0.5806}$ &	$7.4^{+11.0}_{-4.5}\times10^{-6}$ &	$0.85^{+0.30}_{-0.27}$ &	$1.3^{+0.5}_{-0.7}$ &	$154/192$    \\   [2pt]
      \hline\noalign{\vskip2pt}
      2004/6/2 &	N &	$-1.1<$ &	$\dagger$ &		$<0.20$ &			$...$ &			$71/82$    \\   [2pt]
      &	E &	$0.25^{+0.65}_{-0.58}$ &	$1.0^{+1.7}_{-0.6}\times10^{-5}$ &	$0.55^{+0.27}_{-0.24}$ &	$0.9^{+0.8}_{-0.5}$ &	$125/175$    \\   [2pt]
      &	S &	$<5.2$ & 	$2.2^{+5249.7}_{-2.1}\times10^{-7}$ &	$<0.22$ &			$...$ &			$73/79$    \\   [2pt]
      &	W &	$-0.17^{+0.75}_{-0.39}$ &	$7.6^{+10.8}_{-4.6}\times10^{-6}$ &	$1.55^{+0.37}_{-0.33}$ &	$1.9^{+1.0}_{-0.6}$ &	$154/179$    \\   [2pt]
      \hline\noalign{\vskip2pt}
      2004/11/28 &	N &	$-0.16^{+2.29}_{-1.65}$ &	$1.1^{+32.1}_{-1.0}\times10^{-6}$ &	$<0.22$ &			$...$ &			$86/108$    \\   [2pt]
      &	E &	$-0.96^{+0.73}_{-0.67}$ &	$1.2^{+2.6}_{-0.8}\times10^{-6}$ &	$1.56^{+0.36}_{-0.32}$ &	$2.1^{+1.3}_{-0.7}$ &	$160/165$    \\   [2pt]
      &	S &	$0.29^{+5.50}_{-3.17}$ &	$1.5^{+1260.5}_{-1.4}\times10^{-6}$ &	$<0.00$ &			$...$ &			$54/78$    \\   [2pt]
      &	W &	$1.2^{+0.6}_{-0.6}$ &	$3.0^{+7.3}_{-1.5}\times10^{-5}$ &	$0.58^{+0.26}_{-0.23}$ &	$1.3^{+1.2}_{-0.7}$ &	$141/166$    \\   [2pt]
      \hline\noalign{\vskip2pt}
      2010/12/17 &	N &	$...$ &	$\dagger$ &	$<0.10$ &	$...$ &	$196/244$\\   [2pt]
      &	E &	$-0.41^{+0.24}_{-0.24}$ &	$4.0^{+1.7}_{-1.2}\times10^{-6}$ &	$1.31^{+0.16}_{-0.15}$ &	$1.5^{+0.4}_{-0.2}$ &	$341/325$    \\   [2pt]
      &	S &	$2.5<$ &	$\dagger$ &	$0.18^{+0.10}_{-0.09}$ &	$...$ &		$255/245$    \\   [2pt]
      &	W &	$1.7^{+0.3}_{-0.3}$ &	$6.1^{+3.5}_{-2.2}\times10^{-5}$ &	$0.43^{+0.11}_{-0.11}$ &	$1.6^{+0.7}_{-0.6}$ &	$276/299$    \\   [2pt]
      \hline\noalign{\vskip2pt}
      2010/12/24 &	N &	$...$ &	$\dagger$ &	$<0.20$ &	$...$ &	$107/102$    \\   [2pt]
      &	E &	$-0.41^{+0.24}_{-0.24}$ &	$3.9^{+4.3}_{-2.0}\times10^{-6}$ &	$1.20^{+0.33}_{-0.30}$ &	$1.4^{+0.8}_{-0.5}$ &	$235/240$    \\   [2pt]
      &	S &	$...$ &		$\dagger$ &				$<0.19$ &	$...$ &	$71/97$    \\   [2pt]
      &	W &	$1.2^{+0.6}_{-0.5}$ &	$3.4^{+4.9}_{-1.7}\times10^{-5}$ &	$0.29^{+0.22}_{-0.18}$ &	$0.5^{+1.0}_{-0.3}$ &	$175/200$    \\   [2pt]
      \hline\noalign{\vskip2pt}
      2022/7/22 &	N &	$1.8^{+3.3}_{-2.2}$ &	$2.8^{+158.7}_{-2.7}\times10^{-5}$ &	$0.31^{+0.31}_{-0.26}$ &	$...$ &			$54/65$	    \\   [2pt]
	  &	E &	$1.3^{+1.2}_{-1.0}$ &	$3.5^{+8.8}_{-2.7}\times10^{-5}$ &	$<0.29$ &			$0.8^{+2.1}_{-0.8}$ &	$65/107$    \\   [2pt]	
	  &	S &	$1.5^{+3.5}_{-2.6}$ &	$1.3^{+138.5}_{-1.2}\times10^{-5}$ &	$0.34^{+0.31}_{-0.26}$ &	$...$ &			$59/64$	    \\   [2pt]
	  &	W &	$-0.21^{+0.83}_{-0.81}$ &	$5.3^{+13.8}_{-3.9}\times10^{-6}$ &	$1.53^{+0.47}_{-0.41}$ &	$1.9^{+1.5}_{-0.8}$ &	$108/153$    \\   [2pt]
      \hline\noalign{\vskip2pt}
      2022/7/28 &	N &	$-0.25^{+2.93}_{-1.88}$ &	$9.2^{+390.2}_{-9.0}\times10^{-7}$ &	$0.62^{+0.42}_{-0.35}$ &	$3.0^{+37.0}_{-2.4}$ &	$38/66$	    \\   [2pt]
	  &	E &	$1.8^{+1.3}_{-0.9}$ &	$1.1^{+4.3}_{-0.9}\times10^{-4}$ &	$0.79^{+0.43}_{-0.36}$ &	$2.6^{+4.0}_{-1.7}$ &	$69/100$    \\   [2pt]	
	  &	S &	$0.30^{+5.06}_{-1.83}$ &	$9.8^{+2188.8}_{-9.7}\times10^{-6}$ &	$<0.31$ &			$...$ &			$26/40$    \\   [2pt]	
	  &	W &	$-0.96^{+0.57}_{-0.92}$ &	$2.0^{+4.6}_{-1.1}\times10^{-6}$ &	$1.49^{+0.53}_{-0.46}$ &	$1.1^{+0.9}_{-0.5}$ &	$106/136$    \\   [2pt]
      \hline\noalign{\vskip2pt}
    \end{tabular}}\label{tab:all_param}
\begin{tabnote}
    \footnotemark[$*$] Columns: (1) Observation start date. (2) Circumnuclear region name. (3)-(4) Photon index and normalization of the power-law component. For observations where both the upper and lower limits of the photon index could not be determined, or where only the lower limit was determined, the normalization value is not reported and indicated by $\dagger$. This indicates that the normalization value can be extremely large due to the unconstrained photon index.
    (5) Flux of the 6.4\>keV neutral iron K$\alpha$ line for each CNR in units of $10^{-13}$ erg cm$^{-2}$ s$^{-1}$.  
    (6) Equivalent width of the 6.4\>keV neutral iron K$\alpha$ line. EW values are reported only in cases where both the continuum and iron line intensities are constrained by a 90\% confidence contour for two interesting parameters.
    (7) C-statistic value over degrees of freedom. \\
\end{tabnote}
\end{table*}



\begin{figure*}[ht]
 \begin{center}
  \includegraphics[width=14cm]{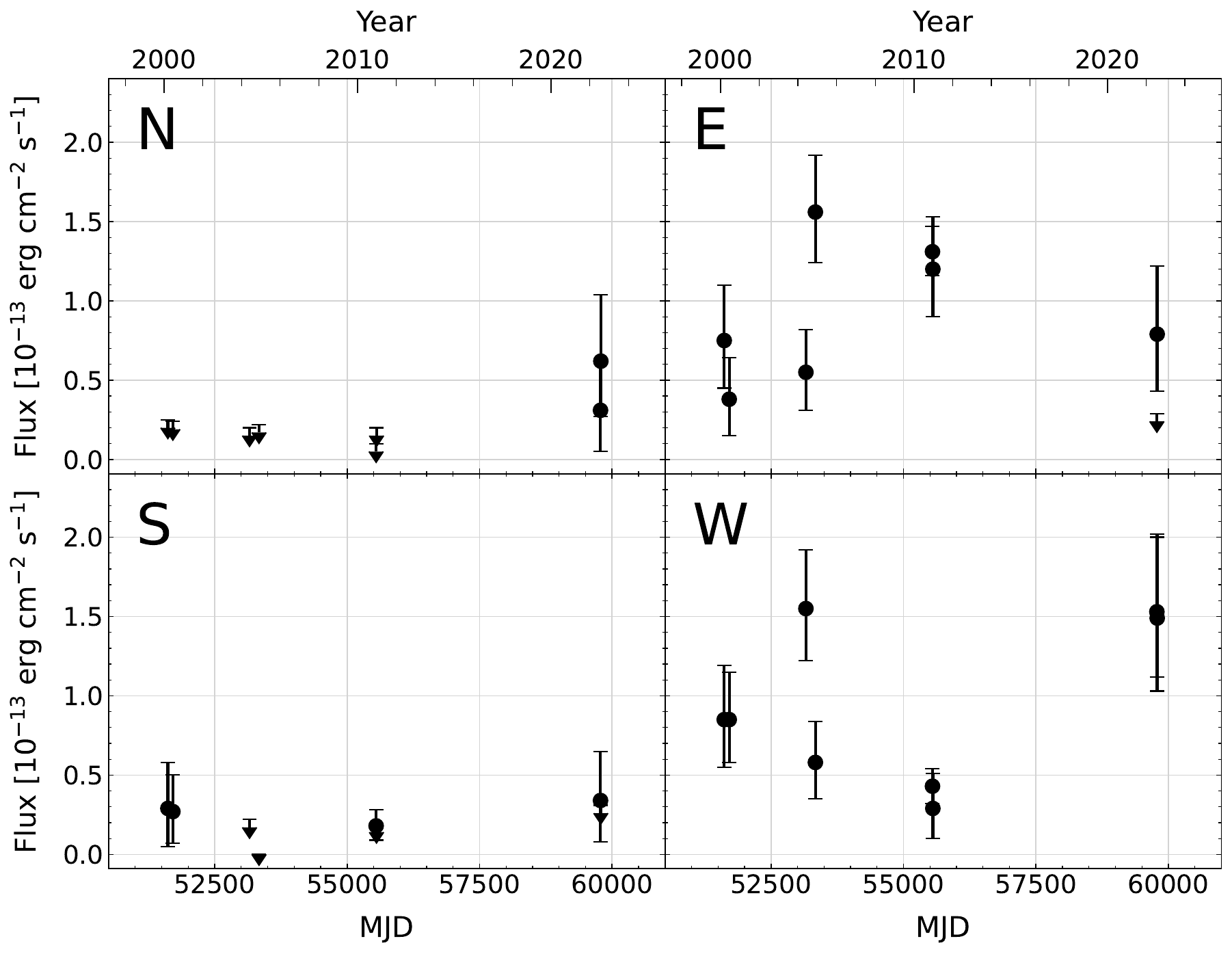} 
 \end{center}
\caption{ Flux of the 6.4\>keV neutral iron emission line from in-situ emissions specific to each CNR. The vertical axis indicates the emission line flux, and the horizontal axis indicates the observation period in Modified Julian Date (MJD). Data points for which only upper limits are available are indicated by inverted triangles.
}\label{fig:Fe_flux}
\end{figure*}


\subsection{Spectral analyses of circumnuclear regions}\label{ssec:cnrspec}

To constrain the fluxes of neutral iron-K$\alpha$ emission in the previously defined four fan-like circumnuclear regions (CNRs), we extracted the spectra from the respective regions and analyzed them while taking account of the contaminating light from the nucleus. 
A background spectrum in each observation was estimated from an annulus of 0\farcm8--1\farcm1 centered at the nucleus. 
We limited the spectral energy band to 3.0--8.0\>keV, sufficient to constrain the underlying continuum as well as the iron-K$\alpha$ emission line at 6.4\>keV.   
The spectra were fitted by considering a power-law model for continuum emission and four Gaussian functions to reproduce iron emission lines: the Fe-K$\alpha$ line at 6.40\>keV, the Fe-K$\beta$ line at 7.06\>keV, the Fe XXV He$\alpha$ line at 6.68\>keV, and the Fe XXVI Ly$\alpha$ line at 6.97\>keV. 
The line energies were fixed at their appropriate values, and all line widths were fixed at $\sigma = 0\>\mathrm{km\>{s}^{-1}}$ as the spectral resolution is poor. 
In addition, we took account of the contaminating light from the central region by adopting and fixing the model determined in sub-subsection~\ref{sssec:nuspec}. 
The response file for the in-situ emission and the contamination from the center was generated using the CIAO tools $\mathtt{specextract}$ and $\mathtt{arfcorr}$.

We were able to reproduce all 32 spectra (eight epochs $\times$ four regions) well by considering the in-situ emission and the contaminating light (see figure~\ref{fig:allcnrspec} in the Appendix for all fitting results).   
Representatively, figure~\ref{fig:cnrspec} shows the fitting results for the spectra on 2004 June 2 and suggests good agreement. Particularly in the regions E and W, the in-situ emission, consisting of prominent iron emission and slightly inverted continuum with photon indices of $-1$ to $0$, is more significant than the contaminating one. 
Table~\ref{tab:all_param} summarizes the constrained fluxes of the iron-K$\alpha$ emission. The regions E and W reached luminosities of up to $L_{\mathrm{Fe-K\alpha}}=3.1^{+0.7}_{-0.6}\>\times10^{38}\>\mathrm{erg\>s^{-1}}$ and $3.1\pm0.7\>\times10^{38}\>\mathrm{erg\>s^{-1}}$, respectively.
Figure~\ref{fig:Fe_flux} shows the light curves for the respective regions.
While we found insignificant or faint emission with fluxes less than $\approx 2\times10^{-14}$ erg cm$^{-2}$ s$^{-1}$ in the regions N and S, the brighter emission was significantly observed in the regions E and W in almost all epochs. 
To statistically examine whether the iron-K$\alpha$ emission varied across the observations, we fitted a constant model to each light curve in the chi-squared method and calculated the null hypothesis probability. 
In the fits, the upper limit data were excluded for simplicity. 
We then confirmed significant variations, particularly for the regions E and W, with the derived probabilities of less than 1\%.

In addition to the line flux, we calculated equivalent widths (EWs), useful in discussing the origin of the iron-K$\alpha$ emission, when the emission was the brightest in each of the regions E and W. 
We estimated the errors in the EWs while considering that the errors in the line flux and the flux of the underlying power-law emission could correlate with each other. 
For each of the regions E and W, we created a 90\% confidence contour between the two relevant parameters with $\mathtt{steppar}$ command in XSPEC and estimated the allowed EW range. 
The resultant EWs were $EW_{\mathrm{Fe-K\alpha}}=2.1^{+1.3}_{-0.7}\>\mathrm{keV}$ in the region E on 2004 November 28 and $EW_{\mathrm{Fe-K\alpha}}=1.9^{+1.0}_{-0.6}\>\mathrm{keV}$ in the region W on 2004 June 2.

\section{Discussion}\label{sec:discussion}

\subsection{Collisional ionization or photoionization as the origin of iron K$\alpha$ emission}\label{ssec:ionization}

In the E and W regions, we detected the significant spatially extended iron-K$\alpha$ emission (section~\ref{ssec:cnrspec}), 
and, from a physical point of view, such fluorescent emission can, in principle, originate from either photoionization or collisional ionization. In fact, observationally, iron-K$\alpha$ emission has been found on scales of $\sim 100$ pc from Sgr\,A$^\ast$ in the Galactic Center region, and various researchers have discussed two plausible origins for the fluorescence: ionization by X-rays from Sgr\,A$^\ast$ (e.g., \citealt{Koyama1996,Nobukawa2010}) and ionization by high-energy particles (\citealt{Yusef-Zadeh2007}). Motivated by these studies, we also examine which of these two mechanisms better explains the origin of the iron-K$\alpha$ emission that we detected.
To distinguish ionizing mechanisms, the observed EWs of iron-K$\alpha$ emission ($\sim$\>2\>keV) and the spectral slope of the underlying continuum emission ($\Gamma \sim -1$ to $0$) can be hints. 

As detailed below, the observed EWs and slopes seem to be inconsistent with the collisional-ionization scenario, whereas photoionization seems to be able to explain them. 
\citet{Tatischeff2012} theoretically estimated the observable quantities of Fe-K$\alpha$ emission and continuum X-rays expected from the interaction of accelerated particles with neutral ambient gas. 
Electrons with energies above $\gtrsim$ 7\>keV ionize irons, producing the iron-K$\alpha$ emission together with free-free continuum emission underneath the line. 
The expected EWs are less than 0.5\>keV over the possible ranges of properties of incident electrons and gas with a solar composition (e.g., electron energy distribution, electron minimum energy, and gas density). 
In contrast, accelerated protons with energies above $\gtrsim$ 1\>MeV can produce  Fe-K$\alpha$ emission with EWs of $\gtrsim$ 1\>keV, if abundant protons with energies of 0.1--1\>MeV are simultaneously available. 
Physically, while protons below 1\>MeV 
can ionize irons and produce the iron emission, their energies are insufficient to produce significant free-free emission around 6.4\>keV (see figure~5 of \citeay{Tatischeff2012}). 
Here, a notable point is that, to have a high EW of $\sim$ 2\>keV, protons should have a soft energy distribution, or a power-law distribution with a slope of $\gtrsim 2$, and such a distribution predicts a soft X-ray spectrum with a photon index of $>$ 1.5. 
These predictions were not consistent with the observed EWs and slopes. 
In contrast, the remaining photoionization scenario can explain the quantities. 
X-ray photons above $\approx$ 7.1\>keV ionize irons and produce iron-K$\alpha$ emission, while the underlying continuum is the result of photoelectronic absorption and Thomson scattering \citep{Matt1991A&A...247...25M}. 
In this scenario, the EW can be higher than 1\>keV, and also a flat or inverted spectrum $\Gamma < 0$ can be reproduced. 
These predictions are consistent with what we observed. 

We note that, as listed in table~\ref{tab:all_param}, the EW became higher than 1\>keV, yet the power-law index was larger than 0 in some observations. This is apparently inconsistent with 
the photoionization scenario, but still may be able to be explained in the same context.
Given that high EWs with hard indices are found particularly when the iron emission line was faint, the X-ray reflection component would have been weak at the periods. 
In the situations, if a different softer X-ray component remains almost constant, fitting the spectrum with a single power-law continuum yields an apparently larger (i.e., softer) photon index (figure~\ref{fig:allcnrspec} in appendix). 
Then, the continuum level around the iron line is largely reduced, and the EW can get larger than 1\>keV correspondingly.

\subsection{Photoionizing sources}\label{ssec:photoionization}

As discussed in subsection~\ref{ssec:ionization}, photoionization is the plausible physical mechanism responsible for the iron line in the E and W regions, and the AGN is a natural candidate for the ionizing source. For completeness, however, it would be useful to also examine the possible contribution from X-ray binaries.  
By searching for previous studies that calculated the EW of iron-K$\alpha$ emission in X-ray binaries, we find that IGR J17252-3616, which is a high-mass X-ray binary (HMXB), exhibited an EW of $\approx$ 2.7\>keV (\citeay{Aftab2019}). 
In addition, we find that, among the low-mass X-ray binaries (LMXBs), GX 1+4 showed an EW of $\approx$ 2.1\>keV (\citeay{Rea2005}). 
Although these EWs are similar to those found for the regions E and W ($\sim\>2.0\>\mathrm{keV}$; subsection~\ref{ssec:cnrspec}), their luminosities are much lower than those we measured ($\sim10^{38}\>\mathrm{erg\>s^{-1}}$); the iron-line luminosities of IGR J17252-3616 and GX 1+4 are $9.4\times10^{32}\>\mathrm{erg\>s^{-1}}$ and $2.7\times10^{34}\>\mathrm{erg\>s^{-1}}$, respectively. 
Therefore, a single X-ray binary would not be the responsible source.
The AGN is eventually left as the candidate, and its primary X-ray luminosity is high enough to reproduce the observed iron-K$\alpha$ luminosity. 
By referring to \citet{Sunyaev1998}, we estimate an expected luminosity to be $\sim10^{38}\>\mathrm{erg\>s^{-1}}$ for a reasonable distance to the gas cloud (40 pc), 
solid angle ($0.01\>\mathrm{sr}$ assumed from the typical size of molecular clouds),
density ($10^{3}\>\mathrm{cm^{-3}}$), and AGN X-ray luminosity ($3\times10^{42}\>\mathrm{erg\>s^{-1}}$ based on \cite{Arevalo2014}). 
The result suggests that the AGN is the probable source responsible for the iron-K$\alpha$ emission, and the time variability may reflect the change in the AGN X-ray emission. 
If correct, our discovery may be the first identification of a time-varying X-ray AGN echo in an extra-galactic SMBH system, whereas the galactic one due to Sgr\,A$^\ast$ has been proposed (e.g., \citeay{Koyama1996}).

\subsection{Size of the cloud illuminated by AGN}\label{ssec:cloudsize}

Under the AGN echo scenario, we roughly estimate the sizes of clouds based on the time scale of flux variation in the iron emission ($\delta t_{\rm Fe}$). The significant variation is confirmed in subsection~\ref{ssec:cnrspec}. 
An upper limit for the illuminated cloud size can be inferred from $\delta t_{\rm Fe}/2\times c$, where $c$ is the speed of light, and we adopt the FWHM of the iron-line time variation for $\delta t_{\rm Fe}$. 
By inspecting figure~\ref{fig:Fe_flux}, we adopt $\delta t_{\rm Fe} \sim $ 18 yrs between 2004 and 2022  and $\sim$ 4 yrs between 2000 and 2004 for the regions E and W, respectively. 
Consequently, the upper limits are estimated to be $\sim$ 6 pc for the region E, and $\sim$ 1 pc for the region W. 
These constraints, particularly for the region W, are better than the canonical Chandra resolution of 0\farcs5 ($\sim$10 pc).

To place more accurate constraints on the sizes of the emitting clouds, 
it is important to understand the relative geometry among the clouds, the AGN, and the observer. 
For the understanding, we need to know in which directions X-ray photons with energies around the iron K edge ($\sim 7.1$ keV) can propagate from the nucleus without being heavily absorbed by the torus. 
Also, it is necessary to where gas capable of reflecting the photons is located. 
While gas distributions of Circinus have been revealed from the torus scale to 
the host-galaxy one at different wavelengths (e.g., \citeay{Freeman1977}; \citeay{Elmouttie1998}; \citeay{Greenhill2003}; \citeay{Tristram2014}; \citeay{Stalevski2019}; \citeay{Isbell2022}; \citeay{Ursini2023}), 
for the torus geometry, we refer to \citet{Uematsu2021}, who put constraints on it by X-ray observations. 
The authors infer an inclination of $\sim 80^\circ$ and a torus half-opening angle of $\sim 80^\circ$ for lines of sight with $N_{\rm H} < 10^{24}\>\mathrm{cm^{-2}}$. 
In this configuration, the AGN can illuminate a wide solid angle in the X-ray band. 
Regarding the three-dimensional distribution of gas on a scale of $\sim 100$ pc, we refer to the ALMA CO($J$=3--2) observation presented in \citet{Izumi2018}. 
Their results indicate a geometrically thin gas distribution and an inclination angle of $\sim 65^\circ$, such that the eastern side corresponds to the near side of the galaxy. 
Combining these constraints, we infer that the cloud responsible for the iron-K$\alpha$ emission in the E region is located in front of the AGN at an angle of $\sim 65^\circ$ from the plane of the sky passing through the AGN. In contrast, those in the W region lie behind the AGN by a similar angle. 
Under these assumptions, we estimate the upper limits of the cloud sizes in the E and W regions to be $\sim 9$ pc and $\sim 0.3$ pc, respectively.

The estimated sizes are comparable to those of molecular clumps, which have hydrogen molecular gas densities of $\sim 10^{3-4}$  cm$^{-3}$ (e.g., \citeay{Minamidani2011}; \citeay{Fujii2014}). Thus, hydrogen column densities seen from the AGN can be of order of $10^{22-23}$ cm$^{-2}$, which would be sufficient to produce iron-K$\alpha$ emission and be consistent with the AGN-irradiation scenario. 

As a side note, one might think that the iron line image (figure~\ref{fig:Fe_map}) may appear more extended than the upper limits obtained above. There are two possible explanations for the apparent discrepancy. First, figure~\ref{fig:Fe_map} shows the iron line image divided by the point-source image, and because of the division, extended emission at larger radii may be enhanced. 
As a result, the iron line distribution can look broader than it actually is. As a simple test, we performed MARX simulations for the E region assuming an iron-K$\alpha$ source with ${\rm FWHM} \sim 9$ pc and confirmed that the resulting image is consistent with the observed one. Second, if the emitting gas consists of multiple clumps that are distributed over a wide area but illuminated nearly simultaneously, the apparent discrepancy may emerge.


\begin{figure*}[ht]
 \begin{center}
  \includegraphics[width=16cm]{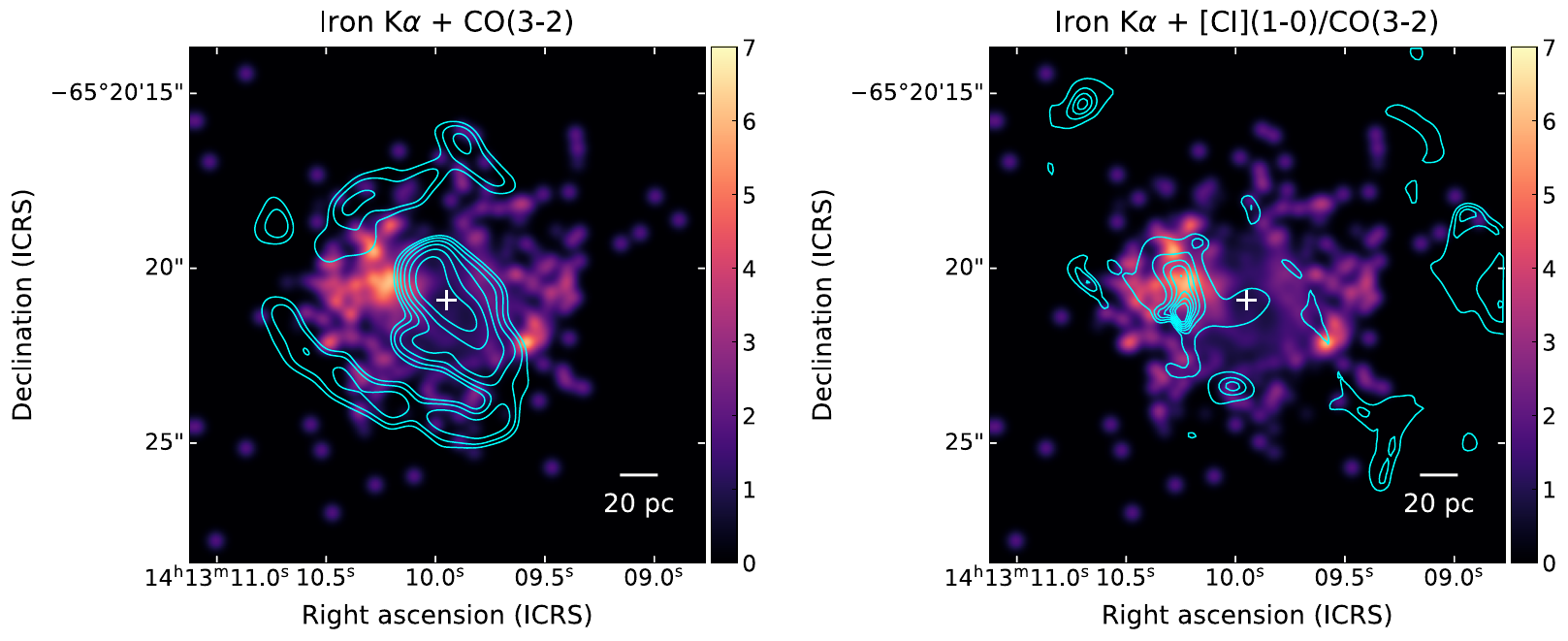} 
 \end{center}
\caption{ Left panel: Iron K$\alpha$ emission map on 2010 December 17 (same as in figure~\ref{fig:Fe_map}) and contours of the CO($J$=3--2) integrated density ($5\times2^{(n-1)/2}\sigma$ with $n=$ 13, ..., and 18, where $1\sigma=0.038\>\mathrm{Jy\>beam^{-1}\>km\>s^{-1}}$) created by \citet{Izumi2018}. 
Right panel: Iron emission map on 2010 December 17 (same as in the left panel) and the [C\,I]($J$=1--0) to CO($J$=3--2) integrated intensity ratio map created by \citet{Izumi2018}.  Contours represent the [C\,I]($J$=1--0)/CO($J$=3--2) ratio, ranging from 0.6 to 3.0 in steps of 0.4, within the region where CO($J$=3--2) emission is significant above 40$\sigma$. 
The AGN position is indicated by a white cross.
}\label{fig:Fe_CO_CICO}
\end{figure*}


\subsection{Comparison with cold molecular gas distribution}\label{ssec:mol}

Our results have revealed the sites where the AGN X-ray emission irradiates the interstellar matter, and it is interesting to see what kinds of impacts the X-ray radiation has on the interstellar matter. 
We compare the iron line map, created in subsection~\ref{ssec:ironmap}, 
with the distribution of cold ($T\sim\>10\>\mathrm{K}$) molecular gas observed by the Atacama Large Millimeter/submillimeter Array (ALMA).
We used the map of CO($J$=3--2) emission, created by \citet{Izumi2018}. 
Figure~\ref{fig:Fe_CO_CICO} shows the comparison, revealing that the region E with bright Fe-K$\alpha$ emission coincides with the region of weaker CO emission. 
Furthermore, by focusing on [C\,I]($J$=1--0)/CO($J$=3--2) shown by \citet{Izumi2018}, we notice that 
a high [C\,I]($J$=1--0)/CO($J$=3--2) ratio was achieved in that region. 
These results might be consistent with the X-ray-dominated region (XDR) model (e.g., \citeay{Maloney96}). 
X-ray photons penetrate into gas, while producing high-energy photo-electrons and destroying molecules. Thus, the atomic-to-molecular gas ratio can be enhanced. 
For further investigation, ALMA follow-up observations of molecular and atomic lines at resolutions sufficient to resolve the gas clouds suggested by the time variability would be important. 
We also mention an alternative scenario that molecular gas may be heated to a high temperature ($T\sim\>1000\>\mathrm{K}$) instead of dissociation, having made it difficult to observe molecular gas using ALMA. 
Thus, for example, JWST achieving a high spatial resolution in the infrared, where hot hydrogen lines appear, would be also important.


\section{Conclusion}\label{sec:conc}

We investigated the spatially extended and time-variable neutral iron K$\alpha$ emission in Circinus using eight Chandra observations spanning over two decades from 2000 to 2022. 
By simulating and subtracting the nuclear point-source emission, we extracted in-situ iron emission especially in the eastern and western sectors, approximately 20–60 pc from the nucleus. Significant variability in the line flux was detected across the two decades of the observations.

The spectral fits revealed large iron-K$\alpha$ EWs up to $\sim$\>2\>keV and flat or inverted, underlying continua (photon index $\Gamma \lesssim 0$) in the extended emission regions. 
These spectral characteristics are inconsistent with collisional ionization by electrons or protons, but are naturally explained by photoionization by hard X-ray photons emitted from the AGN. 
Thus, the observed variability may reflect time-varying AGN X-ray emission, and we might identify the first extragalactic AGN X-ray echo. 
From the timescales of variability, we constrained the sizes of the reprocessing clouds to be $\lesssim$ 6 pc (region E) and $\lesssim$ 1 pc (region W). 
If we furthermore consider the possible relative positions of each cloud to the AGN and the observer, the sizes of the clouds in the E and W regions are estimated to be $\lesssim$ 9 pc and $\lesssim$ 0.3 pc, respectively.
These values are smaller than the canonical resolution of Chandra, which is achievable for Circinus. 

Comparison with the ALMA molecular line maps showed that the iron-bright regions correspond to CO-deficient zones. The zones are possibly shaped by X-ray-driven chemistry or heating and thus are not fully traceable by cold molecular gas alone. 
Further investigations using ALMA and JWST would give us additional insights into the irradiated region.


\begin{ack}
We thank the anonymous referee for their valuable comments and suggestions, which helped improve the quality of this paper. 
Also, we thank Professor Nagamine for his fruitful discussion.
This research has made use of data obtained from the Chandra Data Archive.
We used data analysis software of the application packages CIAO and HEASoft provided by the Chandra X-ray Center (CXC) and the High Energy Astrophysics Science Archive Research Center (HEASARC), respectively.
\end{ack}

\section*{Funding}
 This work is supported by the Japan Science and Technology Agency (JST) Support for Pioneering Research Initiated by the Next Generation (SPRING), Grant Number JPMJSP2138 (AM) and the Japan Society for the Promotion of Science (JSPS) Grants-in-Aid for Scientiﬁc Research (KAKENHI), Grant Numbers 23K13153, 24K00673 (TK), 22K18277, 22H00128 (HO) and 23H00128 (HM).

\section*{Appendix. All spectra of circumnuclear regions}\label{app:allspec}

\begin{figure*}[t]
 \begin{center}
  \includegraphics[width=12.5cm]{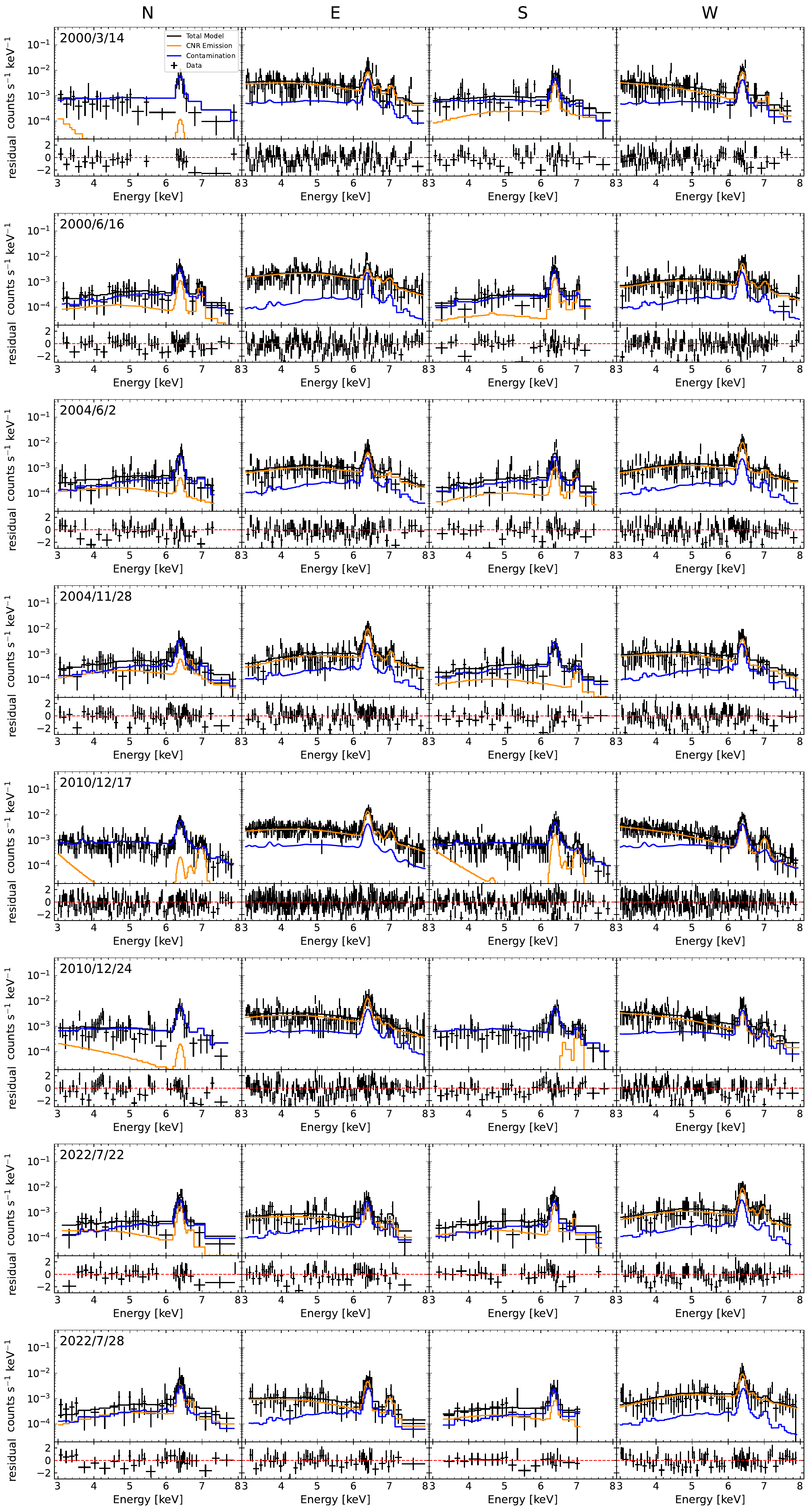} 
 \end{center}
\caption{Spectra and model components of the CNRs in all observations; in-situ emission from each CNR (orange), contamination from the central region (blue), and their combined contribution (black).
}\label{fig:allcnrspec}
\end{figure*}

Figure~\ref{fig:allcnrspec} shows all spectra and models of four CNRs for each observation.


\bibliography{Circinus_paper/ref}

\begin{thebibliography}{}
\expandafter\ifx\csname natexlab\endcsname\relax\def\natexlab#1{#1}\fi
\providecommand{\url}[1]{\href{#1}{#1}}
\providecommand{\dodoi}[1]{doi:~\href{http://doi.org/#1}{\nolinkurl{#1}}}
\providecommand{\doeprint}[1]{\href{http://ascl.net/#1}{\nolinkurl{http://ascl.net/#1}}}
\providecommand{\doarXiv}[1]{\href{https://arxiv.org/abs/#1}{\nolinkurl{https://arxiv.org/abs/#1}}}

\bibitem[{{Aftab} {et~al.}(2019){Aftab}, {Paul}, \& {Kretschmar}}]{Aftab2019}
{Aftab}, N., {Paul}, B., \& {Kretschmar}, P. 2019, \apj, 243, 29, \dodoi{10.3847/1538-4365/ab2a77}

\bibitem[{{Alonso-Herrero} {et~al.}(2020){Alonso-Herrero}, {Pereira-Santaella}, {Rigopoulou}, {Garc{\'\i}a-Bernete}, {Garc{\'\i}a-Burillo}, {Dom{\'\i}nguez-Fern{\'a}ndez}, {Combes}, {Davies}, {D{\'\i}az-Santos}, {Esparza-Arredondo}, {Gonz{\'a}lez-Mart{\'\i}n}, {Hern{\'a}n-Caballero}, {Hicks}, {H{\"o}nig}, {Levenson}, {Ramos Almeida}, {Roche}, \& {Rosario}}]{Alonso-Herrero2020}
{Alonso-Herrero}, A., {Pereira-Santaella}, M., {Rigopoulou}, D., {et~al.} 2020, \aap, 639, A43, \dodoi{10.1051/0004-6361/202037642}

\bibitem[{{Ar{\'e}valo} {et~al.}(2014){Ar{\'e}valo}, {Bauer}, {Puccetti}, {Walton}, {Koss}, {Boggs}, {Brandt}, {Brightman}, {Christensen}, {Comastri}, {Craig}, {Fuerst}, {Gandhi}, {Grefenstette}, {Hailey}, {Harrison}, {Luo}, {Madejski}, {Madsen}, {Marinucci}, {Matt}, {Saez}, {Stern}, {Stuhlinger}, {Treister}, {Urry}, \& {Zhang}}]{Arevalo2014}
{Ar{\'e}valo}, P., {Bauer}, F.~E., {Puccetti}, S., {et~al.} 2014, \apj, 791, 81, \dodoi{10.1088/0004-637X/791/2/81}

\bibitem[{{Arnaud}(1996)}]{Arnaud1996}
{Arnaud}, K.~A. 1996, in Astronomical Society of the Pacific Conference Series, Vol. 101, Astronomical Data Analysis Software and Systems V, ed. G.~H. {Jacoby} \& J.~{Barnes}, 17

\bibitem[{{Bauer} {et~al.}(2015){Bauer}, {Ar{\'e}valo}, {Walton}, {Koss}, {Puccetti}, {Gandhi}, {Stern}, {Alexander}, {Balokovi{\'c}}, {Boggs}, {Brandt}, {Brightman}, {Christensen}, {Comastri}, {Craig}, {Del Moro}, {Hailey}, {Harrison}, {Hickox}, {Luo}, {Markwardt}, {Marinucci}, {Matt}, {Rigby}, {Rivers}, {Saez}, {Treister}, {Urry}, \& {Zhang}}]{Bauer2015}
{Bauer}, F.~E., {Ar{\'e}valo}, P., {Walton}, D.~J., {et~al.} 2015, \apj, 812, 116, \dodoi{10.1088/0004-637X/812/2/116}

\bibitem[{{Canizares} {et~al.}(2005){Canizares}, {Davis}, {Dewey}, {Flanagan}, {Galton}, {Huenemoerder}, {Ishibashi}, {Markert}, {Marshall}, {McGuirk}, {Schattenburg}, {Schulz}, {Smith}, \& {Wise}}]{Canizares05}
{Canizares}, C.~R., {Davis}, J.~E., {Dewey}, D., {et~al.} 2005, \pasp, 117, 1144, \dodoi{10.1086/432898}

\bibitem[{{Davis} {et~al.}(2012){Davis}, {Bautz}, {Dewey}, {Heilmann}, {Houck}, {Huenemoerder}, {Marshall}, {Nowak}, {Schattenburg}, {Schulz}, \& {Smith}}]{Davis2012}
{Davis}, J.~E., {Bautz}, M.~W., {Dewey}, D., {et~al.} 2012, in Society of Photo-Optical Instrumentation Engineers (SPIE) Conference Series, Vol. 8443, Space Telescopes and Instrumentation 2012: Ultraviolet to Gamma Ray, ed. T.~{Takahashi}, S.~S. {Murray}, \& J.-W.~A. {den Herder}, 84431A, \dodoi{10.1117/12.926937}

\bibitem[{{Elmouttie} {et~al.}(1998){Elmouttie}, {Krause}, {Haynes}, \& {Jones}}]{Elmouttie1998}
{Elmouttie}, M., {Krause}, M., {Haynes}, R.~F., \& {Jones}, K.~L. 1998, \mnras, 300, 1119, \dodoi{10.1046/j.1365-8711.1998.02002.x}

\bibitem[{{Fabbiano} {et~al.}(2017){Fabbiano}, {Elvis}, {Paggi}, {Karovska}, {Maksym}, {Raymond}, {Risaliti}, \& {Wang}}]{Fabbiano2017}
{Fabbiano}, G., {Elvis}, M., {Paggi}, A., {et~al.} 2017, \apjl, 842, L4, \dodoi{10.3847/2041-8213/aa7551}

\bibitem[{{Fabbiano} {et~al.}(2018){Fabbiano}, {Paggi}, {Karovska}, {Elvis}, {Maksym}, {Risaliti}, \& {Wang}}]{Fabbiano2018}
{Fabbiano}, G., {Paggi}, A., {Karovska}, M., {et~al.} 2018, \apj, 855, 131, \dodoi{10.3847/1538-4357/aab1f4}

\bibitem[{{Feruglio} {et~al.}(2020){Feruglio}, {Fabbiano}, {Bischetti}, {Elvis}, {Travascio}, \& {Fiore}}]{Feruglio2020}
{Feruglio}, C., {Fabbiano}, G., {Bischetti}, M., {et~al.} 2020, \apj, 890, 29, \dodoi{10.3847/1538-4357/ab67bd}

\bibitem[{{Freeman} {et~al.}(1977){Freeman}, {Karlsson}, {Lynga}, {Burrell}, {van Woerden}, {Goss}, \& {Mebold}}]{Freeman1977}
{Freeman}, K.~C., {Karlsson}, B., {Lynga}, G., {et~al.} 1977, \aap, 55, 445

\bibitem[{{Fujii} {et~al.}(2014){Fujii}, {Minamidani}, {Mizuno}, {Onishi}, {Kawamura}, {Muller}, {Dawson}, {Tatematsu}, {Hasegawa}, {Tosaki}, {Miura}, {Muraoka}, {Sakai}, {Tsukagoshi}, {Tanaka}, {Ezawa}, \& {Fukui}}]{Fujii2014}
{Fujii}, K., {Minamidani}, T., {Mizuno}, N., {et~al.} 2014, \apj, 796, 123, \dodoi{10.1088/0004-637X/796/2/123}

\bibitem[{{Garmire} {et~al.}(2003){Garmire}, {Bautz}, {Ford}, {Nousek}, \& {Ricker}}]{Garmire03}
{Garmire}, G.~P., {Bautz}, M.~W., {Ford}, P.~G., {Nousek}, J.~A., \& {Ricker}, George~R., J. 2003, in Society of Photo-Optical Instrumentation Engineers (SPIE) Conference Series, Vol. 4851, \procspie, ed. J.~E. {Truemper} \& H.~D. {Tananbaum}, 28--44, \dodoi{10.1117/12.461599}

\bibitem[{{Greenhill} {et~al.}(2003){Greenhill}, {Booth}, {Ellingsen}, {Herrnstein}, {Jauncey}, {McCulloch}, {Moran}, {Norris}, {Reynolds}, \& {Tzioumis}}]{Greenhill2003}
{Greenhill}, L.~J., {Booth}, R.~S., {Ellingsen}, S.~P., {et~al.} 2003, \apj, 590, 162, \dodoi{10.1086/374862}

\bibitem[{{Isbell} {et~al.}(2022){Isbell}, {Meisenheimer}, {Pott}, {Stalevski}, {Tristram}, {Sanchez-Bermudez}, {Hofmann}, {G{\'a}mez Rosas}, {Jaffe}, {Burtscher}, {Leftley}, {Petrov}, {Lopez}, {Henning}, {Weigelt}, {Allouche}, {Berio}, {Bettonvil}, {Cruzalebes}, {Dominik}, {Heininger}, {Hogerheijde}, {Lagarde}, {Lehmitz}, {Matter}, {Meilland}, {Millour}, {Robbe-Dubois}, {Schertl}, {van Boekel}, {Varga}, \& {Woillez}}]{Isbell2022}
{Isbell}, J.~W., {Meisenheimer}, K., {Pott}, J.-U., {et~al.} 2022, \aap, 663, A35, \dodoi{10.1051/0004-6361/202243271}

\bibitem[{{Izumi} {et~al.}(2018){Izumi}, {Wada}, {Fukushige}, {Hamamura}, \& {Kohno}}]{Izumi2018}
{Izumi}, T., {Wada}, K., {Fukushige}, R., {Hamamura}, S., \& {Kohno}, K. 2018, \apj, 867, 48, \dodoi{10.3847/1538-4357/aae20b}

\bibitem[{{Izumi} {et~al.}(2023){Izumi}, {Wada}, {Imanishi}, {Nakanishi}, {Kohno}, {Kudoh}, {Kawamuro}, {Baba}, {Matsumoto}, {Fujita}, \& {Tristram}}]{Izumi2023}
{Izumi}, T., {Wada}, K., {Imanishi}, M., {et~al.} 2023, Science, 382, 554, \dodoi{10.1126/science.adf0569}

\bibitem[{{Jones} {et~al.}(2021){Jones}, {Parker}, {Fabbiano}, {Elvis}, {Maksym}, {Paggi}, {Ma}, {Karovska}, {Siemiginowska}, \& {Wang}}]{Jones2021}
{Jones}, M.~L., {Parker}, K., {Fabbiano}, G., {et~al.} 2021, \apj, 910, 19, \dodoi{10.3847/1538-4357/abe128}

\bibitem[{{Kawamuro} {et~al.}(2019){Kawamuro}, {Izumi}, \& {Imanishi}}]{Kawamuro2019}
{Kawamuro}, T., {Izumi}, T., \& {Imanishi}, M. 2019, \pasj, 71, 68, \dodoi{10.1093/pasj/psz045}

\bibitem[{{Kawamuro} {et~al.}(2020){Kawamuro}, {Izumi}, {Onishi}, {Imanishi}, {Nguyen}, \& {Baba}}]{Kawamuro2020}
{Kawamuro}, T., {Izumi}, T., {Onishi}, K., {et~al.} 2020, \apj, 895, 135, \dodoi{10.3847/1538-4357/ab8b62}

\bibitem[{{Kawamuro} {et~al.}(2021){Kawamuro}, {Ricci}, {Izumi}, {Imanishi}, {Baba}, {Nguyen}, \& {Onishi}}]{Kawamuro2021}
{Kawamuro}, T., {Ricci}, C., {Izumi}, T., {et~al.} 2021, \apjs, 257, 64, \dodoi{10.3847/1538-4365/ac2891}

\bibitem[{{Koyama}(2018)}]{Koyama2018}
{Koyama}, K. 2018, \pasj, 70, R1, \dodoi{10.1093/pasj/psx084}

\bibitem[{{Koyama} {et~al.}(1996){Koyama}, {Maeda}, {Sonobe}, {Takeshima}, {Tanaka}, \& {Yamauchi}}]{Koyama1996}
{Koyama}, K., {Maeda}, Y., {Sonobe}, T., {et~al.} 1996, \pasj, 48, 249, \dodoi{10.1093/pasj/48.2.249}

\bibitem[{{Li} {et~al.}(2003){Li}, {Kastner}, {Prigozhin}, \& {Schulz}}]{Li2003}
{Li}, J., {Kastner}, J.~H., {Prigozhin}, G.~Y., \& {Schulz}, N.~S. 2003, \apj, 590, 586, \dodoi{10.1086/374967}

\bibitem[{{Li} {et~al.}(2004){Li}, {Kastner}, {Prigozhin}, {Schulz}, {Feigelson}, \& {Getman}}]{Li2004}
{Li}, J., {Kastner}, J.~H., {Prigozhin}, G.~Y., {et~al.} 2004, \apj, 610, 1204, \dodoi{10.1086/421866}

\bibitem[{{Ma} {et~al.}(2023){Ma}, {Elvis}, {Fabbiano}, {Balokovi{\'c}}, {Maksym}, \& {Risaliti}}]{Ma2023}
{Ma}, J., {Elvis}, M., {Fabbiano}, G., {et~al.} 2023, \apj, 948, 61, \dodoi{10.3847/1538-4357/acba8d}

\bibitem[{{Maloney} {et~al.}(1996){Maloney}, {Hollenbach}, \& {Tielens}}]{Maloney96}
{Maloney}, P.~R., {Hollenbach}, D.~J., \& {Tielens}, A.~G.~G.~M. 1996, \apj, 466, 561, \dodoi{10.1086/177532}

\bibitem[{{Marinucci} {et~al.}(2017){Marinucci}, {Bianchi}, {Fabbiano}, {Matt}, {Risaliti}, {Nardini}, \& {Wang}}]{Marinucci2017}
{Marinucci}, A., {Bianchi}, S., {Fabbiano}, G., {et~al.} 2017, \mnras, 470, 4039, \dodoi{10.1093/mnras/stx1551}

\bibitem[{{Marinucci} {et~al.}(2013){Marinucci}, {Miniutti}, {Bianchi}, {Matt}, \& {Risaliti}}]{Marinucci2013}
{Marinucci}, A., {Miniutti}, G., {Bianchi}, S., {Matt}, G., \& {Risaliti}, G. 2013, \mnras, 436, 2500, \dodoi{10.1093/mnras/stt1759}

\bibitem[{{Marinucci} {et~al.}(2012){Marinucci}, {Risaliti}, {Wang}, {Nardini}, {Elvis}, {Fabbiano}, {Bianchi}, \& {Matt}}]{Marinucci2012}
{Marinucci}, A., {Risaliti}, G., {Wang}, J., {et~al.} 2012, \mnras, 423, L6, \dodoi{10.1111/j.1745-3933.2012.01232.x}

\bibitem[{{Matt} {et~al.}(1991){Matt}, {Perola}, \& {Piro}}]{Matt1991A&A...247...25M}
{Matt}, G., {Perola}, G.~C., \& {Piro}, L. 1991, \aap, 247, 25

\bibitem[{{Matt} {et~al.}(1999){Matt}, {Guainazzi}, {Maiolino}, {Molendi}, {Perola}, {Antonelli}, {Bassani}, {Brandt}, {Fabian}, {Fiore}, {Iwasawa}, {Malaguti}, {Marconi}, \& {Poutanen}}]{Matt1999}
{Matt}, G., {Guainazzi}, M., {Maiolino}, R., {et~al.} 1999, \aap, 341, L39.
\newblock \doarXiv{astro-ph/9811301}

\bibitem[{{Minamidani} {et~al.}(2011){Minamidani}, {Tanaka}, {Mizuno}, {Mizuno}, {Kawamura}, {Onishi}, {Hasegawa}, {Tatematsu}, {Takekoshi}, {Sorai}, {Moribe}, {Torii}, {Sakai}, {Muraoka}, {Tanaka}, {Ezawa}, {Kohno}, {Kim}, {Rubio}, \& {Fukui}}]{Minamidani2011}
{Minamidani}, T., {Tanaka}, T., {Mizuno}, Y., {et~al.} 2011, \aj, 141, 73, \dodoi{10.1088/0004-6256/141/3/73}

\bibitem[{{Nakata} {et~al.}(2021){Nakata}, {Hayashida}, {Noda}, {Yoneyama}, {Matsumoto}, \& {Imanishi}}]{Nakata2021}
{Nakata}, R., {Hayashida}, K., {Noda}, H., {et~al.} 2021, \pasj, 73, 338, \dodoi{10.1093/pasj/psab001}

\bibitem[{{Nobukawa} {et~al.}(2010){Nobukawa}, {Koyama}, {Tsuru}, {Ryu}, \& {Tatischeff}}]{Nobukawa2010}
{Nobukawa}, M., {Koyama}, K., {Tsuru}, T.~G., {Ryu}, S.~G., \& {Tatischeff}, V. 2010, \pasj, 62, 423, \dodoi{10.1093/pasj/62.2.423}

\bibitem[{{Rea} {et~al.}(2005){Rea}, {Stella}, {Israel}, {Matt}, {Zane}, {Segreto}, {Oosterbroek}, \& {Orlandini}}]{Rea2005}
{Rea}, N., {Stella}, L., {Israel}, G.~L., {et~al.} 2005, \mnras, 364, 1229, \dodoi{10.1111/j.1365-2966.2005.09646.x}

\bibitem[{{Sambruna} {et~al.}(2001){Sambruna}, {Netzer}, {Kaspi}, {Brandt}, {Chartas}, {Garmire}, {Nousek}, \& {Weaver}}]{Sambruna2001}
{Sambruna}, R.~M., {Netzer}, H., {Kaspi}, S., {et~al.} 2001, \apjl, 546, L13, \dodoi{10.1086/318068}

\bibitem[{{Soldi} {et~al.}(2005){Soldi}, {Beckmann}, {Bassani}, {Courvoisier}, {Landi}, {Malizia}, {Dean}, {de Rosa}, {Fabian}, \& {Walter}}]{Soldi2005}
{Soldi}, S., {Beckmann}, V., {Bassani}, L., {et~al.} 2005, \aap, 444, 431, \dodoi{10.1051/0004-6361:20053875}

\bibitem[{{Stalevski} {et~al.}(2019){Stalevski}, {Tristram}, \& {Asmus}}]{Stalevski2019}
{Stalevski}, M., {Tristram}, K. R.~W., \& {Asmus}, D. 2019, \mnras, 484, 3334, \dodoi{10.1093/mnras/stz220}

\bibitem[{{Sunyaev} \& {Churazov}(1998)}]{Sunyaev1998}
{Sunyaev}, R., \& {Churazov}, E. 1998, \mnras, 297, 1279, \dodoi{10.1046/j.1365-8711.1998.01684.x}

\bibitem[{{Tatischeff} {et~al.}(2012){Tatischeff}, {Decourchelle}, \& {Maurin}}]{Tatischeff2012}
{Tatischeff}, V., {Decourchelle}, A., \& {Maurin}, G. 2012, \aap, 546, A88, \dodoi{10.1051/0004-6361/201219016}

\bibitem[{{Tristram} {et~al.}(2014){Tristram}, {Burtscher}, {Jaffe}, {Meisenheimer}, {H{\"o}nig}, {Kishimoto}, {Schartmann}, \& {Weigelt}}]{Tristram2014}
{Tristram}, K. R.~W., {Burtscher}, L., {Jaffe}, W., {et~al.} 2014, \aap, 563, A82, \dodoi{10.1051/0004-6361/201322698}

\bibitem[{{Tsunemi} {et~al.}(2001){Tsunemi}, {Mori}, {Miyata}, {Baluta}, {Burrows}, {Garmire}, \& {Chartas}}]{Tsunemi2001}
{Tsunemi}, H., {Mori}, K., {Miyata}, E., {et~al.} 2001, \apj, 554, 496, \dodoi{10.1086/321338}

\bibitem[{{Uematsu} {et~al.}(2021){Uematsu}, {Ueda}, {Tanimoto}, {Kawamuro}, {Setoguchi}, {Ogawa}, {Yamada}, \& {Odaka}}]{Uematsu2021}
{Uematsu}, R., {Ueda}, Y., {Tanimoto}, A., {et~al.} 2021, \apj, 913, 17, \dodoi{10.3847/1538-4357/abf0a2}

\bibitem[{{Ursini} {et~al.}(2023){Ursini}, {Marinucci}, {Matt}, {Bianchi}, {Marin}, {Marshall}, {Middei}, {Poutanen}, {Rogantini}, {De Rosa}, {Di Gesu}, {Garc{\'\i}a}, {Ingram}, {Kim}, {Krawczynski}, {Puccetti}, {Soffitta}, {Svoboda}, {Tombesi}, {Weisskopf}, {Barnouin}, {Perri}, {Podgorny}, {Ratheesh}, {Zaino}, {Agudo}, {Antonelli}, {Bachetti}, {Baldini}, {Baumgartner}, {Bellazzini}, {Bongiorno}, {Bonino}, {Brez}, {Bucciantini}, {Capitanio}, {Castellano}, {Cavazzuti}, {Ciprini}, {Costa}, {Del Monte}, {Di Lalla}, {Di Marco}, {Donnarumma}, {Doroshenko}, {Dovciak}, {Ehlert}, {Enoto}, {Evangelista}, {Fabiani}, {Ferrazzoli}, {Gunji}, {Heyl}, {Iwakiri}, {Jorstad}, {Karas}, {Kitaguchi}, {Kolodziejczak}, {La Monaca}, {Latronico}, {Liodakis}, {Maldera}, {Manfreda}, {Marscher}, {Mitsuishi}, {Mizuno}, {Muleri}, {Ng}, {O'Dell}, {Omodei}, {Oppedisano}, {Papitto}, {Pavlov}, {Peirson}, {Pesce-Rollins}, {Petrucci}, {Pilia}, {Possenti}, {Ramsey}, {Rankin}, {Romani}, {Sgr{\`o}}, {Slane}, {Spandre}, {Tamagawa}, {Tavecchio},
  {Taverna}, {Tawara}, {Tennant}, {Thomas}, {Trois}, {Tsygankov}, {Turolla}, {Vink}, {Wu}, {Xie}, \& {Zane}}]{Ursini2023}
{Ursini}, F., {Marinucci}, A., {Matt}, G., {et~al.} 2023, \mnras, 519, 50, \dodoi{10.1093/mnras/stac3189}

\bibitem[{{Yang} {et~al.}(2009){Yang}, {Wilson}, {Matt}, {Terashima}, \& {Greenhill}}]{Yang2009}
{Yang}, Y., {Wilson}, A.~S., {Matt}, G., {Terashima}, Y., \& {Greenhill}, L.~J. 2009, \apj, 691, 131, \dodoi{10.1088/0004-637X/691/1/131}

\bibitem[{{Young} {et~al.}(2001){Young}, {Wilson}, \& {Shopbell}}]{Young2001}
{Young}, A.~J., {Wilson}, A.~S., \& {Shopbell}, P.~L. 2001, \apj, 556, 6, \dodoi{10.1086/321561}

\bibitem[{{Yusef-Zadeh} {et~al.}(2007){Yusef-Zadeh}, {Muno}, {Wardle}, \& {Lis}}]{Yusef-Zadeh2007}
{Yusef-Zadeh}, F., {Muno}, M., {Wardle}, M., \& {Lis}, D.~C. 2007, \apj, 656, 847, \dodoi{10.1086/510663}

\end{thebibliography}
\bibliographystyle{aasjournal}

\end{document}